%% file: Lebed_LomCom.tex
\documentclass[
preprint,
 amsmath,amssymb,
 aps,
showkeys,
showpacs
]{revtex4-1}

\usepackage{comment}
\usepackage{graphicx}
\usepackage{dcolumn}
\usepackage{bm}

\begin{document}


\title{All Heavy Tetraquarks:
\\The Dynamical Diquark Model
and Other Approaches}

\author{Richard F. Lebed}
 \email{richard.lebed@asu.edu}
\affiliation{%
 Arizona State University\\
 Department of Physics\\ Tempe, Arizona, USA 85287-1504
}%




\date{January, 2022}

\begin{abstract}
The 2020 announcement by LHCb of a narrow structure $X(6900)$ in the
di-$J/\psi$ spectrum---a potential $c\bar c c\bar c$ state---has
opened a new era in hadronic spectroscopy.  In this talk, we briefly
survey theory works preceding this event, examine key features of the
observed spectrum, and then discuss how subsequent theory studies
(including with the author's own dynamical diquark model) have
interpreted these features.  We conclude with proposals for
experiments capable of distinguishing competing interpretations. 
%
\end{abstract}

\maketitle


\section{Introduction} \label{sec:Intro}

This talk nominally discusses all-heavy tetraquarks in general, but
in fact it focuses almost exclusively upon developments related to
the interesting structures reported by LHCb in the di-$J/\psi$
($c\bar c c\bar c$) channel~\cite{Aaij:2020fnh}, particularly a
putative resonance called $X(6900)$.  Early indications for the
observation of possible di-$\Upsilon(1S)$ ($b\bar b b\bar b$) states
above 18~GeV~\cite{Durgut:2018,ANDY:2019bfn} were not supported by
subsequent searches~\cite{LHCb:2018uwm,CMS:2020qwa}.  Furthermore, no
structures have yet been reported in other all-heavy channels, such
as $b \bar b c \bar c$, $b \bar c b \bar c$, {\it etc}.
Nevertheless, quark-flavor universality suggests that any approach
successful for $c\bar c c\bar c$ should be applicable to all of these
sectors.

All-heavy tetraquark production at hadron colliders is believed to
be proceed primarily through gluon-gluon fusion, with both single-
and double-parton scattering being important for nonresonant
$Q\bar Q Q^\prime \! \bar Q^\prime$
production~\cite{Maciula:2020wri}.  Heavy-ion collisions have also
been identified as important potential production sources for such
states~\cite{Zhao:2020nwy,Esposito:2021ptx}.

$X(6900)$ joins a large family (now over 50!) of heavy-quark
tetraquark (and pentaquark) candidates discovered in the past two
decades~\cite{Lebed:2016hpi}, including hidden-charm,
hidden-bottom, and open-charm candidates.  Of these, the Particle
Data Group~\cite{Zyla:2020zbs} identifies 15 as fully established.
Moreover, a naive count through the various flavor sectors,
upon employing any widely discussed substructure (hadronic molecules,
diquark compounds, {\it etc.}), suggests that over 100 more such
exotic states await discovery.

\section{Features of the LHC\lowercase{b} Data} \label{sec:LHCbData}

\begin{figure}[h]
\vspace{-0.25in}
\includegraphics[scale=0.5]{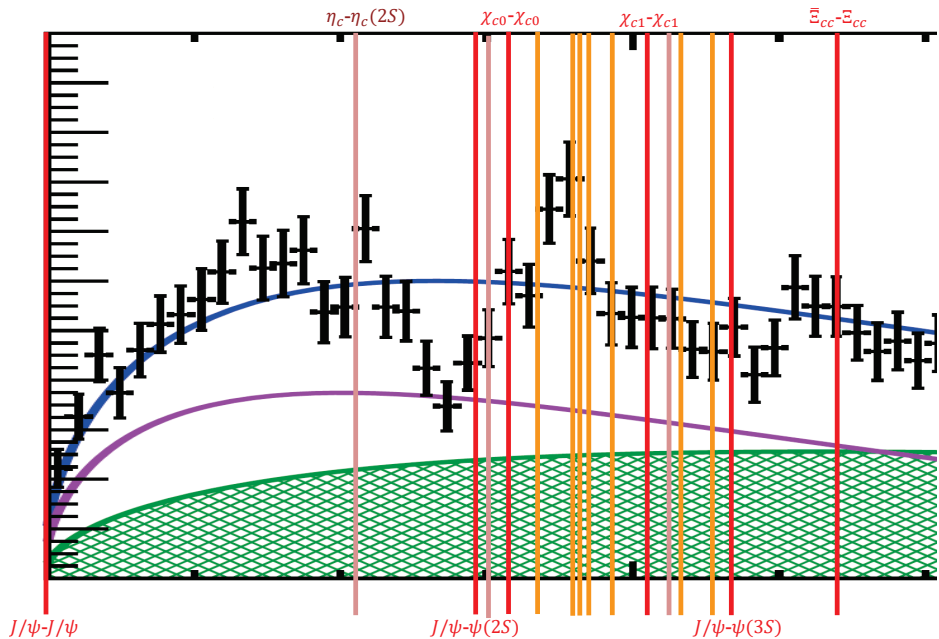}
\vspace{-0.25in}
\caption{\label{fig:LHCbData} Di-$J/\psi$ production data, adapted from
Ref.~\cite{Aaij:2020fnh}.  Ticks on the horizontal axis mark every
200~MeV in the range 6200--7400~MeV\@.  Vertical lines denote
thresholds for di-hadron states containing $c\bar c c\bar c$, red and orange
indicating pairs of hadrons with the same $c\bar c$ spin structure as
$J/\psi$.  The curves indicate fits to components of the nonresonant
contribution carried out in Ref.~\cite{Aaij:2020fnh}.}
\end{figure}

The LHCb data contains not only the $X(6900)$, but also several other
features of interest.  Figure~\ref{fig:LHCbData} adapts a portion of
Fig.~3(a) of Ref.~\cite{Aaij:2020fnh}, focusing upon the region of
interest that starts at the di-$J/\psi$ threshold
$\simeq \! 6200$~MeV and extends slightly beyond the
$c\bar c c\bar c$ open-flavor threshold at
$2m_{\Xi_{cc}} \! \simeq \! 7240$~MeV\@.  One immediately notes not
only the prominent peak near 6900~MeV, but also a broad excess above
background around 6400--6500~MeV, a sharp dip near 6750~MeV, and a
rapid transition from a dip to an excess just above 7200~MeV\@.

$X(6900)$ is the only obvious peak, and although it lies about
700~MeV above $2m_{J/\psi}$, it is likely not wider
($\alt \! 200$~MeV) than the $\rho$, and potentially is much
narrower.  In fact, a $c\bar c c\bar c$ state behaving as a
``traditional'' di-hadron molecule would bind through exchanging
conventional charmonium states; since the expected molecular binding
energies are expected to be $\le \! O(10~{\rm MeV})$, the exchanged
charmonium that binds $X(6900)$ would lie far off mass shell.
Typical $c\bar c$ mean charge radii from potential models are about
0.35~fm for $1S$ states, and about twice that for $1P$ and $2S$
states.  One concludes that $J/\psi$ exchange, in particular, would
be very short-ranged, implying a molecular state in which all quarks
occupy the same volume.  Alternately, one may consider
non-traditional di-hadron molecules in which the $J/\psi$ pair are
bound through Pomeron (multi-gluon) exchanges~\cite{Gong:2020bmg} or
through the exchange of soft gluons that hadronize into light-meson
($\pi , K$) exchange quanta~\cite{Dong:2021lkh}.

Of particular note are the allowed $J^{PC}$ quantum numbers for an
identical $J/\psi \, (1^{--})$ pair: In the $S$ wave one can have
$0^{++}$ or $2^{++}$, while the allowed $P$ waves are $0^{-+}$,
$1^{-+}$, and $2^{-+}$.  These $J^{PC}$ values refer both to
di-$J/\psi$ resonances and the background.  Alternately, if one
considers the di-$J/\psi$ system only in terms of identifcal $cc$ and
$\bar c \bar c$ fermion pairs, then in their respective $S$ waves,
only (color-$\bar{\bf 3}$, spin-1) and (color-${\bf 6}$, spin-0)
combinations are allowed.

\section{Theoretical Studies} \label{sec:Theory}

\vspace{-0.125in}
\subsection{Prior to the LHCb Observation}
\vspace{-0.125in}

Remarkably, the first discussion of a di-$J/\psi$ bound
resonance~\cite{Iwasaki:1975pv} followed the 1974 dis\-covery of
charmonium~\cite{E598:1974sol,SLAC-SP-017:1974ind} by only one year.
Afterwards, however, only 1 further paper (by the same author as
Ref.~\cite{Iwasaki:1975pv}) followed in the 1970s, 5 in the 1980s, 1
in the 1990s, and 3 in the 2000s.  The reason for this dearth of
effort appears to stem from the realization that di-$J/\psi$ states
were likely to lie above the di-$J/\psi$ threshold.  As one of the
early works stated, ``All the states are unbounded and consequently
rather uninteresting.''

2010 saw first physics from the LHC, and it was very quickly
appreciated that the collider would produce a great deal of
$gg \! \to \! J/\psi \, J/\psi$, which would be straightforward for
LHCb to reconstruct via $J/\psi \! \to \mu^+ \mu^-$.  One theory
collaboration analyzed this effect and predicted (using a diquark
model) the spectrum of $c\bar c c\bar c$ states that could be
produced~\cite{Berezhnoy:2011xy,Berezhnoy:2011xn}.  While one might
have expected a subsequent flurry of work using other theoretical
approaches, no further papers studying the $c\bar c c\bar c$ system
appeared until 2016, with 12 more papers (perhaps anticipating the
trove of LHC Run~2 data) appearing from 2016 to the middle of 2020\@.

\vspace{-0.25in}
\subsection{Subsequent to the LHCb Observation}
\vspace{-0.125in}

The LHCb observations were announced in a talk on
16~June~2020~\cite{An:2020}, and the collaboration's preprint was
posted on 30~June~2020~\cite{Aaij:2020fnh}.  In that 2-week period
alone, 8 more theory papers appeared.  Since then (as of the time of
writing), 58 additional papers discussing either the production or
spectroscopy of these states have been produced.  The length limit
imposed upon this document precludes even the mere listing of these
references, let alone discussing all of them in any detail.  Instead,
we examine the features in the LHCb data, with an eye to their
interpretation as proposed in a select few papers.

A wide variety of theoretical approaches have been employed to
understand the data, including: string junction models; quark models
with chromomagnetic interactions; quark potential models; chiral
quark models; diquark models; effective theories with light-meson
exchanges; threshold effects with coupled charmonium channels;
threshold effects plus compact tetraquarks; QCD sum rules; lattice
simulations; Regge phenomenology including Pomeron exchange;
holography; spin-chain (Bethe Ansatz) algebraic methods; and the
Bethe-Salpeter approach.  $X(6900)$ has even been proposed to be a
Higgs-like boson~\cite{Zhu:2020snb}.

Are there any solid conclusions or points of consensus uniting these
analyses?  First, not many authors dispute that $X(6900)$ seems to be
a genuine resonance, even when including the effects due to the
presence of multiple thresholds (see Fig.~\ref{fig:LHCbData}) that
might explain other $c\bar c c\bar c$ structure, as considered in,
{\it e.g.}, Refs.~\cite{Dong:2020nwy,Guo:2020pvt}.  However, other
authors ({\it e.g.}, Ref.~\cite{Wang:2020wrp}) suggest that even
$X(6900)$ itself might be generated by the $\chi_{c0}$-$\chi_{c1}$
threshold.

Second, virtually all models (back to the very first
papers~\cite{Iwasaki:1975pv}) predict ground-state ($1S$)
$c\bar c c\bar c$ resonances to lie much lower than $X(6900)$,
typically from 6000-6400~MeV\@.  So then, is $X(6900)$ a
$P \! = \! -$, $1P$ state ({\it e.g.},
Ref.~\cite{Sonnenschein:2020nwn}), or a $P \! = \! +$, $2S$ state
({\it e.g.}, Refs.~\cite{Giron:2020wpx,Karliner:2020dta})?

The broad structure around 6400-6500~MeV lies at the upper limit of
where models predict $1S$ ground states to occur ({\it e.g.},
Refs.~\cite{Yang:2020wkh,Zhao:2020zjh}).  In LHCb's
Model~I~\cite{Aaij:2020fnh}, for example, the structure is treated as
a superposition of (at least) two resonances, irrespective of quantum
numbers.  Indeed, what is meant by ``$1S$'', which suggests a 2-body
description?  Since ``traditional''  molecules are problematic for
$c\bar c c\bar c$, and no good thresholds lie in the 6400-6500~MeV
range (Fig.~\ref{fig:LHCbData}), then the diquark
$(cc)_{\bar{\bf 3}} (\bar c \bar c)_{\bf 3}$ structure seems most
natural for identifying possible quantum numbers.  But not everyone
agrees!  Noting that ${\bf 6}$-$\bar{\bf 6}$ attraction is stronger
than $\bar{\bf 3}$-${\bf 3}$ (despite quark repulsion in a {\bf 6}
diquark), Ref.~\cite{Deng:2020iqw} finds that the ground states mix
both configurations, while the $\bar{\bf 3}$-${\bf 3}$ configuration
dominates excited states.

The dip around 6750~MeV suggests destructive interference with
$X(6900)$.  LHCb's Model~II~\cite{Aaij:2020fnh} posits interference
between the broad 6400-6500~MeV structure and a second resonance.  A
$\chi_{c0}$-$\chi_{c0}$ threshold effect provides an alternate
explanation~\cite{Karliner:2020dta}.  Furthermore, if $X(6900)$ is
$2S$ ($P \! = \! +$), then $1P$ states ($P \! = \! -$) are expected
near 6750~MeV~\cite{Giron:2020wpx}, although $P \! = \! \pm$
configurations do not interfere with each other; again, measuring
parities is crucial.

The structure near 7200~MeV may be associated with open-flavor decays
of $c\bar c c\bar c$, which first appear at the
$\bar \Xi_{cc}$-$\Xi_{cc}$ [$(ccu)(\bar c\bar c \bar u)$]  threshold
7242.4(1.0)~MeV\@.  Likely, no narrow $c\bar c c\bar c$ structures
occur above this point.  This is the point in diquark models where
the color flux tube breaks~\cite{Giron:2020wpx}, or in holographic
models where new string junctions become
possible~\cite{Sonnenschein:2020nwn}.

How many states are expected in a diquark model?  If both
$\bar{\bf 3}$ and ${\bf 6}$ diquarks are allowed, one finds a
{\em lot\/}~\cite{Bedolla:2019zwg}: 17 with $C \! = \! +$ and
$J \! \le \! 2$ are predicted to lie below the
$\bar \Xi_{cc}$-$\Xi_{cc}$ threshold.  If one adopts a minimal
ansatz of allowing only $\bar{\bf 3}$ diquarks, then only about half
as many states occur.  Further taking quark spin couplings to be
large only within diquarks (a defining property of the dynamical
diquark model~\cite{Lebed:2017min,Giron:2020wpx}), then all $S$-wave
multiplets consist of 3 degenerate states: $0^{++}$, $1^{+-}$, and
$2^{++}$, and all $P$-wave multiplets consist of 7 states (3 with
$C \! = \! +$) that follow an equal-spacing mass rule if tensor
couplings are neglected.

\section{Some Parting Proposals} \label{sec:Thoughts}
\vspace{-0.125in}

Most obviously, we desperately need $J^P$ information in order to
disentangle the di-$J/\psi$ spectrum.  An excellent suggestion
({\it e.g.}, Refs.~\cite{Richard:2020hdw,Cao:2020gul}) is to look at
the $J/\psi$-$\psi(2S)$ spectrum, even though its threshold is
700~MeV higher, and $\psi(2S)$ production is lower than that of
$J/\psi$.  Note in this regard that BESIII sees distinct $Y$ exotics
decaying to $J/\psi$ or to $\psi(2S)$.

The $gg$ production of di-$J/\psi$ is $C \! = \! +$; is there much
$ggg$ ($C \! = \! -$) production?  If so, one could find the $1^{+-}$
resonance via $J/\psi$-$\eta_c$, although $\eta_c$ is harder to
reconstruct [but note that B.R.($\eta_c \! \to \! p\bar p$) =
$1.45 \! \times \! 10^{-3}$].  Alternately, $J/\psi$-$\chi_{cJ}$ also
has $C \! = \! -$, but less phase space. 

And don't forget about $c\bar c b\bar b$ and $b\bar b b\bar b$
production!  $c\bar c b\bar b$ (via $J/\psi$-$\Upsilon$) should
produce many more resonances, by evading the identical-fermion
constraint.   Such experiments provide important tests of
quark-flavor universality.

\paragraph*{Acknowledgments.}
This work was supported by the US National Science Foundation (NSF)
under Grants No.\ PHY-18030912 and PHY-2110278.

\input{Lebed_LomCom.bbl}

\end{document}

%% file: Lebed_LomCom.bbl
%